\documentclass[useAMS]{mn2e}

\usepackage{graphicx}

\title[]{}

\author[]{}

\title[Limiting central density of Dark Matter Haloes] {An upper limit
  to the central density of dark matter haloes from consistency with
  the presence of massive central black holes} \author[X. Hernandez \&
  W. H. Lee] {X. Hernandez and William H.~ Lee \\ Instituto de
  Astronom\'{\i}a, Universidad Nacional Aut\'{o}noma de M\'{e}xico,
  Apartado Postal 70--264 C.P. 04510 M\'exico D.F. M\'exico. \\}

\date{Released 2007 Xxxxx XX}

\pagerange{\pageref{firstpage}--\pageref{lastpage}} \pubyear{2009}

\def\LaTeX{L\kern-.36em\raise.3ex\hbox{a}\kern-.15em

    T\kern-.1667em\lower.7ex\hbox{E}\kern-.125emX}

\begin{document}

\label{firstpage}

\maketitle

\begin{abstract}

We study the growth rates of massive black holes in the centres of
galaxies from accretion of dark matter from their surrounding
haloes. By considering only the accretion due to dark matter
particles on orbits unbound to the central black hole, we obtain a
firm lower limit to the resulting accretion rate. We find that a
runaway accretion regime occurs on a timescale which depends on the
three characteristic parameters of the problem: the initial
mass of the black hole, and the volume density and velocity dispersion
of the dark matter particles in its vicinity. An analytical treatment of the
accretion rate yields results implying that for the
largest black hole masses inferred from QSO studies ($>10^{9}
M_{\odot}$), the runaway regime would be reached on time scales which
are shorter than the lifetimes of the haloes in question for central
dark matter densities in excess of $250 M_{\odot}$~pc$^{-3}$. Since
reaching runaway accretion would strongly 
distort the host dark matter halo, the inferences of QSO black
holes in this mass range lead to an upper limit on the central dark
matter densities of their host haloes of $\rho_{0} < 250 M_{\odot}
$~pc$^{-3}$. This limit  scales inversely with the assumed
central black hole mass. However, thinking of dark matter profiles as
universal across galactic populations, as cosmological studies imply,
we obtain a firm upper limit for the central density of dark matter in
such structures.
\end{abstract}


\section{Introduction} \label{intro}

The question of what is the central density profile of galactic dark
matter haloes has been much debated in the literature over
many years.  Since the work of Navarro et. al (1997), cosmological
N-body simulations have consistently agreed in yielding central
density profiles which can be accurately fitted by functional forms
characterised by centrally divergent density cusps.  Although the
details vary and the innermost slope and radius to which
said fits are to be trusted are still discussed, a broad agreement has
been reached in that cosmologically simulated dark haloes exhibit what
are termed 'cuspy density profiles', e.g. Merritt et al. (2006).

On the other hand, observational inferences of dark halo density
profiles through rotation curve decomposition, have tended to favour
those showing density profiles which tend to constant
values  towards the centre. Recent examples include de Blok et
al. (2008), Kuzio de Naray et al. (2009) and Gebhardt \& Thomas
(2009). The issue is complicated by the necessity to estimate the
dynamical relevance of the baryonic component, a function of the
assumed mass to light ratio, the relevance of observational
uncertainties such as beam smearing, and the 
importance of non circular motions and non centrifugal support of
asymmetric drift and hydrodynamic pressure terms (e.g. Valenzuela et
al. 2007).

An interesting independent clue to the puzzle might come from the
presence of massive black holes in the centre of dark matter
haloes. The relevance of such black holes is well established from
their role as the central engines for quasars and active galaxies
(Rees 1984) as well as in quiescent systems (Kormendy \& Richstone
1995). Although mostly seen as active nuclei in the distant universe,
it is commonly believed that all large galaxies host such objects  
in their centres. Recent empirical determinations of central QSO 
black hole masses have established their existence with masses in the
range $10^{7} - 10^{10} M_{\odot}$ at redshifts beyond $z\simeq 3$
(Kelly et al. 2008, Graham 2008).

We therefore must conclude that the central regions of large dark
haloes have coexisted with massive black holes over most of the
history of the universe. Given the existence of event horizons
associated to black holes, and the assumption of standard cold dark matter
subject only to gravitational interactions, it follows that central
black holes have grown over the history of galactic dark haloes,
through the capture of dark matter particles.

The problem of the growth of single central galactic black holes
through accretion was addressed by Lynden-Bell \& Rees
(1971) and more recently by Gnedin \& Primack (2004) and Zhao et
al. (2002), but only in the accretion of particles on capture orbits. 
While readily acreted, they constitute only a
minor fraction of those available in the distribution function of halo
particles. Further, once absorbed, one has to wait over a
comparatively long halo relaxation timescale for it to be re-populated
with dark matter particles. Here we consider only the accretion of
unbound particles, through the absorption
cross section presented by the black hole through its event horizon,
enhanced by gravitational focusing. In this case, the accretion is
slower at first, but proceeds at an ever increasing rate as the
growing black hole mass leads to an increase in its area, with little
accompanying depletion of the overall dark matter halo distribution
function. This, as accretion does not takes place over a highly
specific fraction of the distribution function, while the black hole
mass constitutes only a fraction of the overall dark halo mass.

We find the process to be
characterised by the onset of a rapid runaway growth phase after
a  critical timescale. This timescale is a function of
the mass of the black hole and the local density of dark matter.
By requiring that the runaway phase does not occur, as then the
swallowing up of the halo by the black hole would seriously distort
the former, we can obtain upper limits to the maximum allowed
density of dark matter at the centres of haloes.





\section{Central Black Hole Growth Rates}\label{analytical}

In the case of cuspy dark matter profiles, the scales over which the
density varies by a significant factor, even in the central regions,
are typically of order $10$~pc or above (Merritt et al. 2006,
Stadel et al. 2009). This is many orders of magnitude greater than the
typical length scales over which the accretion onto the central black
hole of mass $M$ takes place, the Schwarzschild radius, $R_{\rm Sch}=5
\times 10^{-4}{\rm pc} (M/5 \times 10^{9} M_{\odot})$. The range of
scales involved, from $R_{\rm Sch}$ to the scale of the dark matter
halo, make a full simulation of the whole problem unfeasible.  To
first approximation, we will therefore treat the central region of the dark matter
halo over which the black hole finds itself as one of constant density.

The effects of central black holes, mostly binary ones, on the stellar
population of a galaxy have been treated before, e.g Quinlan (1996),
Merritt \& Milosavljevic (2005), Sesana et al. (2007).  The problem
here is different because the total mass of the dark halo, and the
range of radii covered by the dark matter particles are both much
larger than equivalent quantities in the case of bulge stars affected
by central black holes. Since the mass of the black hole is still
several orders of magnitude smaller than the total mass of the dark
halo, we shall treat the presence of the black hole as a perturbation
on the distribution function of the dark matter particles.

In Hernandez \& Lee (2008), we calculated the density response of a
constant density, isothermal dark matter distribution, to the presence
of a point mass $M$. A highly localised cusp appears, with the density
profile changing from $\rho_{0}$ to $\rho_{0}+\rho_{1}(r)$, where

\begin{equation}
\rho_{1}(r)=\frac{G M}{r \sigma^{2}} \rho_{0}.
\end{equation}

In the above, $\sigma$ is the velocity dispersion of the halo
particles, and $r$ is the distance to the mass $M$. In estimating the
growth rate of the central black hole, of initial mass $M_{0}$, we
shall begin by assuming that its presence will elicit a response in
the otherwise unperturbed dark halo particles, as described by
eq.~(1). In Hernandez \& Lee (2008) we showed
through direct comparison with high resolution N-body simulations,
that the analytic expression in eq.~(1) accurately
describes the response of a constant density isothermal region to the
presence of a point mass. A first correction due to relativistic
effects, the substitution of a Newtonian potential for the expression
of Paczy\'{n}ski \& Wiita (1980)\footnote{In the P-W expression a point
mass produces a gravitational field $\Phi(r)=-GM/(r-R_{\rm Sch})$
instead of the usual $\Phi=-GM/r$.}, will result only in the
substitution of $r-R_{\rm Sch}$ for the current $r$ in the denominator of
eq. (1).

We can now estimate the growth rate of the central black hole
dimensionally as:

\begin{equation}
\dot{M}=C_{0} \rho A \sigma,
\end{equation}

where $\rho$, $A$ and $\sigma$ are a characteristic density, area and
velocity for the spherical accretion in question, and $C_{0}$ is a
dimensionless constant which one would expect to be of order
unity. From the preceding discussion regarding the reaction of the
halo to the presence of the black hole, and from the fact that
accretion of unbound particles will occur on crossing the event
horizon at $R=R_{\rm Sch}$, we can estimate $\dot{M}$ from taking
$\rho = \rho_{1}(R_{\rm Sch})$, $A=4 \pi R_{\rm Sch}^{2}$ and $\sigma$
as the velocity dispersion of the dark matter particles in the
halo. We have ignored $\rho_{0}$ in favour of $\rho_{1}$ in the above
considerations, as at distances of order $R_{Sch}$ the latter dominates 
over the former by a factor of order $(c/\sigma)^{2}$. The above yields:

\begin{equation}
\dot{M}=C_{0} 8 \pi \frac{G^{2} M^{2}\rho_{0}}{\sigma c^{2}}.
\end{equation}

Gravitational focusing effectively increases the cross section of the
black hole by trapping particles which would otherwise fail to be
acreted, and enters into the computation of $C_{0}. $In fact, a fully
relativistic calculation for the accretion rate of a black hole
immersed in an isothermal distribution of non-relativistic particles
leads to the result:

\begin{equation}
\dot{M}=16 \left( 6 \pi \right)^{1/2} \frac{G^{2} M^{2}\rho_{0}}{\sigma c^{2}},
\end{equation}

as derived in Shapiro \& Teukolsky (1983), eq.~14.2.26, for particles with positive energies.
Those on bound orbits can be engulfed rapidly in a short initial transient phase, and will
re-appear as the corresponding phase space is re-populated by the
distribution function on the very long relaxation timescales of the
full halo (Shapiro \& Teukolsky 1983).  Here again,
by considering only the accretion rate of unbound particles, we are
confident in having a secure lower limit on the accretion
rate. Comparing with the dimensional analysis of eq.~(3) we see
that the result is exact for $C_{0}=2(6/ \pi)^{1/2}$, a factor of less
than 3. In what follows we shall use the exact result of eq.~(4),
with eq.~(3) serving only in allowing a physical
interpretation of the relativistic result.

Introducing dimensionless quantities $\Sigma=\sigma/c$, $\tau=t/t_{\rm ff}$ 
and ${\mathcal M}=M/M_{\rm J}$, where $t_{\rm ff}= 1/(G \rho_{o})^{1/2}$ is the free
fall timescale of the unperturbed background density,
and $M_{\rm J}=\sigma^{3}/(G^{3/2} \rho_{0}^{1/2})$
is the Jeans mass of the unperturbed halo, we obtain the dimensionless
growth rate:

\begin{equation}
\frac{d{\mathcal M}}{d \tau} = (6 \pi)^{1/2}(4 {\mathcal M} \Sigma)^{2}.
\end{equation}

We note that in cosmological N-body simulations, the distribution
function of dark matter particles exhibits a large
degree of orbital anisotropy and is dominated by highly radial orbits
(e.g. Ascasibar \& Gottlober 2008). The increased fraction of the dark
halo hence available for interaction with the central black hole, and
the reduced angular momentum of the dark matter particles, compared to
the isothermal case, will all tend to yield faster growth
rates than those calculated here. The upper limits on central
density derived below are hence safe upper estimates. More detailed
calculations accounting for an intrinsically cusped dark halo profile
with radially dominated distribution functions would yield even lower
limit densities.

\section{Central Dark Matter Density Limits}

From eq.~(4) we obtain for the black hole mass:

\begin{equation}
M(t)=\frac{M_{0} c^{2} \sigma}{c^{2}\sigma - 16(6 \pi)^{1/2} G^{2}\rho_{0} M_{0} t}.
\end{equation}

There is a strong divergence for $t \rightarrow
t_{\rm div} = c^{2} \sigma /(16(6 \pi)^{1/2} G^{2} \rho_{0} M_{0})$. The time 
for it to appear decreases as the
initial black hole mass rises, as the central dark matter density
increases, and increases as the velocity dispersion of the halo
particles rises. The divergence is so abrupt that the
time it takes for the black hole mass to increase by one order of
magnitude, $T_{10}$, is only 9/10 $t_{\rm div}$. The evolution we
calculate will still be accurate up to $t=T_{10}$, as the total mass
of the dark halo (several times $10^{12} M_{\odot}$ for large galactic
haloes), will still be over an order of magnitude larger than that of
the central black hole, even for the largest initial black hole masses
considered, of a few times $10^{9} M_{\odot}$. We shall therefore define:

\begin{equation}
T_{10} = \left( \frac{9}{10} \right) \frac{c^{2} \sigma}{16 (6 \pi)^{1/2} G^{2} \rho_{0} M_{0}},
\end{equation}
as a characteristic timescale after which the accretion process
results in substantial dynamical alterations to the overall dark
halo. With the same dimensionless quantities as defined previously,
we obtain the corresponding expressions:

\begin{equation}
{\mathcal M}(\tau)=\frac{{\mathcal M}_{0}}{1-C_{1}{\mathcal M}_{0} \Sigma^{2} \tau},
~~~~~~~ \tau_{\rm div}=\left(C_{1} {\mathcal M}_{0} \Sigma^{2}  \right)^{-1},
\end{equation}
where $C_{1}=16(6 \pi)^{1/2}$.
We can now calculate the evolution of eq. (6) for any
value of the central black hole mass. We begin with parameters as
appropriate for the largest inferred QSO central black holes, $M_{0}=5
\times 10^{9} M_{\odot}$ (Kelly et al. 2008, Graham 2008), as this
case  will lead to the most restrictive dark matter density
limits.  Although the dark haloes of QSOs cannot be observationally
inferred, given the scalings observed at low redshift between black
hole masses and galactic properties (e.g. Kormendy \& Richstone 1995,
Gebhardt et al. 2000, Ferrarese \& Merritt 2000, Tremaine et al. 2002,
Gultekin et al. 2009), between black hole masses and halo properties
(e.g. Volonteri et al. 2003, Bolton et al. 2008, Croton 2009, Bandara
et al. 2009) and between galactic
properties and dark halo masses, such as the Tully-Fisher relation
(Tully \& Fisher 1977), it is reasonable to assume that QSOs hosting
the largest inferred black hole masses will be hosted by large
galactic haloes. Consequently, we take a large value of $\sigma=200$
~km/s. Given the usual scaling between $\sigma$ and the flat rotation
curve velocity of a galactic halo of $V_{rot} = 2^{1/2} \sigma$, this choice corresponds to $V_{rot} = 280$~km/s, a
value in the extreme range for any type of galactic system. Notice
that as $t_{\rm div}$ scales with $\sigma$, taking a large value for
this parameter will again result in conservative upper limits on the
final inferred limit central halo densities (see below). Also, given
the 'inside out' and 'downsizing' aspects of current cosmological
structure formation models, (e.g. Naab et al. 2009) the dynamical
stability of the central regions of the most massive galactic systems
over the lifetimes we have assumed appears reasonable.

Figure~1 shows the growth of a central black hole as a function
of time, for our fiducial case with $M_{0} =5 \times 10^{9} M_{\odot}$
and $\sigma=200$~km/s, and a range of values for the assumed central
dark matter density $\rho_{0}$=400, 350, 300, 250, 200, 150 and 100
$M_{\odot}$~pc$^{-3}$ (from top to bottom, respectively). The
case where $T_{10}=10$~Gyr corresponds to the middle curve, where the
central dark matter density is $\rho_{0}=250 M_{\odot}$~pc$^{-3}$. The
mass of the black hole increases by a factor of 10 in 10~Gyr, a
conservative estimate of the lifetime of the systems in question,
namely, dark haloes of QSOs observed at high redshift.
We see that for central dark matter densities above this threshold of
$250 M_{\odot}$~pc$^{-3}$, the mass of the central black hole enters
the runaway accretion regime and diverges on timescales shorter than
the lifetimes of the systems being treated, as given by the three
upper curves in Figure~1. On the
other hand, for values below this threshold,
the growth of the central black hole is of only a factor of order
unity over 10~Gyr, as shown by the three lower curves.

\begin{figure}
\includegraphics[angle=0,scale=0.4]{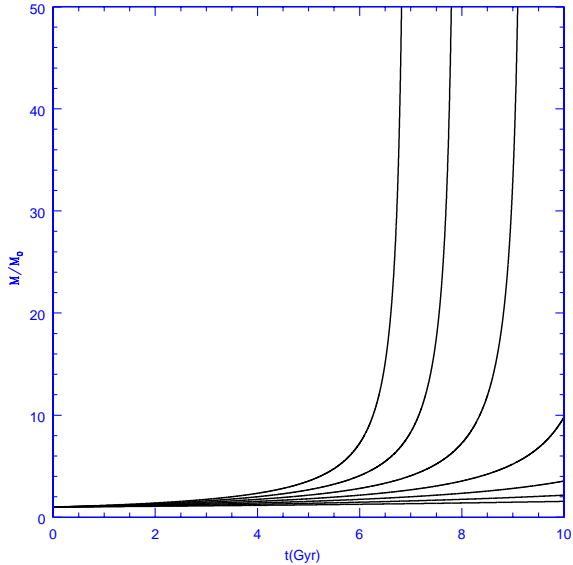}
\caption{Evolution of the mass of a central black hole with initial
mass $M_{0}=5 \times 10^{9} M_{\odot}$ in a dark halo with dark matter
particles of isotropic velocity dispersion $200$~km/s, for
varying central region dark matter density:
$\rho_{0}$=400, 350, 300, 250, 200, 150 and 100
$M_{\odot}$~pc$^{-3}$, top to bottom, respectively.}
\label{fig:picture}
\end{figure}

For this particular initial mass, $M_0=5 \times 10^{9}M_{\odot}$, we
can hence identify $250 M_{\odot}$~pc$^{-3}$ as a maximum central halo
dark matter density above which the inferences of black hole masses in
high redshift QSOs would imply growth rates for the central black
holes resulting in substantial dynamical distortions, leading today
not to quiescent black holes in the centres of normal galaxies, but to
exotic objects dynamically dominated by extreme super massive black
holes. Consistency arguments of this type can be found e.g. in Gnedin 
\& Ostriker (2001), who calibrate the physical parameters of self-interacting
dark matter by requiring that galactic dark haloes should not have evaporated
by now into galaxy cluster dark matter haloes. If we were to take larger 
values for the black hole mass, such as those given by Graham (2008), 
reaching $10^{10} M_{\odot}$, or upwards of $10^{10} M_{\odot}$ reported 
for some objects by Kelly et al. (2008), the threshold density we 
identify would go down by a factor of a few.

In Figure~2 we show the constraints on the central dark halo
densities, $\rho_{M}$, as a function of the assumed central black hole
mass and for a fixed value of the velocity dispersion,
$\sigma=200$~km/s. The curves correspond to various values of $T_{10}$. 
By taking a conservative measure of the lifetimes of the systems in question as
10~Gyr, the region containing the dotted curves above the thick black
line at $T_{10}=10$~Gyr is excluded from consistency arguments, while
the allowed region of parameter space lies in the half plane below it.
The choice of values for the lifetimes of high redshift
QSOs larger than the 10~Gyr previously assumed, shifts the maximum
central dark matter density values downwards onto the various thin
solid curves. The most stringent limits apply to the
highest black hole masses at $M \geq 5 \times
10^{9}M_{\odot}$. However, the expectation of universality for the
cosmological dark matter density profiles leads one to expect these
limits will apply to all dark matter haloes.

\begin{figure}
\includegraphics[angle=0,scale=0.4]{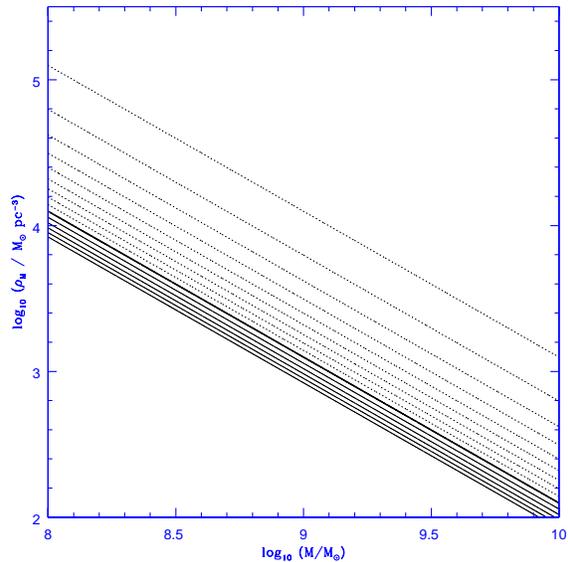}
\caption{Limit central densities as a function of central black hole
masses, for a number of values of $T_{10}$ going from 1 to 15 Gyr,
every Gyr, from top to bottom respectively. The region above the
thick black line at $T_{10}=10$~Gyr, with dotted lines, is excluded
from consistency arguments.}
\label{fig:picture}
\end{figure}

We note that once the mass of the central black hole grows
substantially, processes not included here would begin to
become relevant, and would invalidate the simple physical
hypothesis leading to eq.~(4). Some include: the
adiabatic contraction of the dark halo in response to the
concentration of mass into the central black hole, resulting in
higher central dark matter densities and hence even higher accretion
rates; the accretion of a fraction of the baryons into the central
black hole, which is known to occur; or the enhanced
gravitational focusing of matter of all types into the black hole,
once the approximation of the black hole mass being small compared to
the total halo mass which we are working under begins to break
down. All of these make it reasonable to assume that
the first corrections to eq.~(4) will lead to even larger
accretion rates, hence leaving our conclusions, in terms of limit
densities, unchanged.

Regarding the accretion of baryons and dark halo particles, it has
been proposed that this can be partly responsible for the
appearance of the observed scaling laws between central black hole
masses and bulge and galactic properties, e.g., in the analytical work
of Zhao et al. (2002) and Gnedin \& Primack (2004), the large scale
simulations of Di Matteo et al. (2008), or the luminosity function
consistency arguments in Yu \& Tremaine (2002) and Hopkins et
al. (2007). Indeed, Hennawi \& Ostriker (2002) constrain the velocity
dependence of the interaction cross-section of hypothetical
self-interacting dark matter, by requiring that the accretion of dark
matter onto central black holes leads to the observed $M-\sigma$
relation. Also for the case of self-interacting dark matter, Balberg
\& Shapiro (2002) explore the formation of supermassive black
holes through gravothermal core-collapse of the central regions of
galactic dark haloes.

Comparing the upper limiting central dark matter density of 
$250 M_{\odot}$~pc$^{-3}$ with the dynamically inferred structure of galactic dark
haloes, it is reassuring that when a constant density core is
used to model observations, the inferred central dark matter 
densities always lie below this limit, typically at $\simeq 1
M_{\odot}$~pc$^{-3}$, or below. Recent examples  are given by
Gilmore et al. (2007) for local dwarf spheroidal
galaxies, and de Blok et al. (2008) for late type galaxies.
Hence no conflict appears, in that the runaway accretion regime for
the central black hole will not be reached in 10 Gyr for any 
directly inferred values of the central dark matter density, for any
inferred central black hole masses.

From the point of view of cuspy dark matter haloes, the limits we
derive here establish an inner boundary, exterior to which the
globally fitted centrally divergent dark halo profiles can be
valid. At smaller radii, this solution must be modified to avoid the
divergent black hole growth rates found here. Although our high limit
densities are not reached by current cosmological N-body simulations,
which typically stop at volume densities of order $1
M_{\odot}$~pc$^{-3}$ at their resolution limit, even for the most
recent highest resolution experiments (Cuesta et al. 2008, Stadel et
al. 2009), the logarithmic slopes in these regions are still such that
volume densities above the limits we derive here would be reached
orders of magnitude in the radial coordinate before reaching the scale
of the super massive central black holes, even for central black hole
masses below the upper ranges of these values. Studies of
the origin of cuspy cosmological density profiles
within the secondary infall scenario have traced the cusp to the close
to scale free initial perturbation spectrum (e.g. Williams et al. 2004, 
Salvador-Sol\'{e} et al. 2005, Del Popolo 2009). Such studies explain the steep negative
logarithmic slopes of density profiles from N-body simulations,
predicting them to extend into the very centres. Although these
considerations apply only to cosmological dark matter haloes in the
absence of any central black holes, the inclusion of single central
black holes, currently beyond the reach of fully self consistent
simulations, will result in even steeper central dark matter profiles,
strengthening the consistency arguments made here. A solution
might include dark matter physics not ordinarily
considered, such as self-interacting dark matter 
(e.g. Firmani et al. 2000, de la Macorra 2009), warm dark
matter, or changes to the initial fluctuation spectrum
(e.g. Alam et al. 2002).

\section{Conclusions}\label{ccl}
We study the mass growth rates of central black holes through
accretion of dark matter particles on orbits unbound to the central
black hole. As the black hole mass grows, a runaway
accretion regime ensues. Requiring that no such runaway regime has
been reached over lifetimes of galactic dark haloes of 10~Gyr leads
to the identification of critical upper limits for
the density of dark matter. These limits
scale proportionally to the assumed value of the dark matter velocity dispersion, and
inversely proportionally to the assumed value of the central black hole mass.


For the largest black hole masses inferred by QSO studies of 
$5 \times 10^{9} M_{\odot}$, central region dark matter densities larger than 
$\rho_{0} = 250 M_{\odot}$~pc$^{-3}$ are excluded.
These limits suggest dark halo density structures are characterised by
constant density central regions, rather than divergent cuspy profiles.

\section*{acknowledgements}

The authors acknowledge the constructive criticism of an anonymous 
referee in helping to reach a more clear and complete final version.
Financial support for this work was provided in part through CONACyT
(83254E), and DGAPA-UNAM (PAPIIT IN-113007-3 and IN-114107).

\end{document}